\documentclass[prd,nofootinbib,preprint,superscriptaddress,twocolumn,10pt]{revtex4}
\pdfoutput=1

\usepackage{amsmath,amssymb}
\usepackage{epsfig}
\usepackage{graphicx}
\usepackage[usenames,dvipsnames]{color}
\usepackage{subfigure}
\usepackage{slashed}
\usepackage{comment}
\usepackage[colorlinks,citecolor=blue]{hyperref}
\usepackage{color}

\newcommand{\bmt}{\begin{pmatrix}}
\newcommand{\emt}{\end{pmatrix}}
\newcommand{\ba}{\begin{array}{c}}
\newcommand{\ea}{\end{array}}
\newcommand{\be}{\begin{equation}}
\newcommand{\ee}{\end{equation}}
\newcommand{\bea}{\begin{eqnarray}}
\newcommand{\eea}{\end{eqnarray}}

\newcommand{\bi}{\begin{itemize}}
\newcommand{\ei}{\end{itemize}}

\newcommand{\baz}{\begin{array}{cc}}
\newcommand{\besub}{\begin{subequations}}
\newcommand{\eesub}{\end{subequations}}

\newcommand{\mathsym}[1]{{}}

\newcommand{\bt}{\begin{tabular}}
\newcommand{\et}{\end{tabular}}

\newcommand{\benu}{\begin{enumerate}}
\newcommand{\eenu}{\end{enumerate}}

\def\l{\lambda}

\def\q2 {q^2}

\def\bt{\begin{table}}
\def\et{\end{table}}

\usepackage[utf8]{inputenc}
\usepackage{amsmath,amssymb}
\usepackage{epsfig}
\usepackage{comment}
\usepackage{graphicx}
\usepackage[usenames,dvipsnames]{color}
\usepackage{float}
\usepackage{bbold}
\usepackage[colorlinks,citecolor=blue]{hyperref}
\usepackage{color}
\usepackage{subfigure}

\begin{document}
\title{Baryon asymmetry from dark matter decay}


\author{Debasish Borah}
\email{dborah@iitg.ac.in}
\affiliation{Department of Physics, Indian Institute of Technology Guwahati, Assam 781039, India}

\author{Suruj Jyoti Das}
\email{suruj@iitg.ac.in}
\affiliation{Department of Physics, Indian Institute of Technology Guwahati, Assam 781039, India}

\author{Rishav Roshan}
\email{rishav.roshan@gmail.com}
\affiliation{Department of Physics, Kyungpook National University, Daegu 41566, Korea}
\affiliation{Center for Precision Neutrino Research, Chonnam National University, Gwangju 61186, Korea}

\begin{abstract}
We propose a novel framework where baryon asymmetry can arise due to forbidden decay of dark matter (DM) enabled by finite temperature effects in the early universe. In order to implement it in a realistic setup, we consider the DM to be a singlet Dirac fermion which acquires a dark asymmetry from a scalar field $\Phi$ via Affleck-Dine mechanism. Due to finite-temperature effects, DM can decay in the early universe into leptons and a second Higgs doublet thereby transferring a part of the dark asymmetry into lepton asymmetry with the latter getting converted into baryon asymmetry subsequently via electroweak sphalerons. DM becomes stable below a critical temperature leading to a stable relic. While the scalar field $\Phi$ can play the role of inflaton with specific predictions for inflationary parameters, the setup also remains verifiable via astrophysical as well as laboratory based observations.

\end{abstract}

\maketitle

\noindent
{\bf Introduction:} The matter content of the present universe is dominated by dark matter (DM) with the visible matter comprising only around $20\%$ of it. Additionally, the visible or baryonic matter component is highly asymmetric \cite{Aghanim:2018eyx, Zyla:2020zbs}. While the standard model (SM) can not solve these longstanding puzzles of DM and baryon asymmetry of universe (BAU), several beyond standard model (BSM) proposals have been proposed in the last few decades. Among them, the weakly interacting massive particle (WIMP) paradigm of DM \cite{Kolb:1990vq, Jungman:1995df, Bertone:2004pz, Feng:2010gw, Arcadi:2017kky, Roszkowski:2017nbc} and baryogenesis/leptogenesis \cite{Weinberg:1979bt, Kolb:1979qa, Fukugita:1986hr} have been the most widely studied ones. While the fundamental origin of DM and BAU could be different, the striking similarity in their abundances namely, $\Omega_{\rm DM} \approx 5\,\Omega_{\rm Baryon}$ might be hinting towards a common origin. Such cogenesis mechanisms broadly fall into two categories: one in which DM sector is also asymmetric, known as asymmetric dark matter (ADM) \cite{Nussinov:1985xr, Davoudiasl:2012uw, Petraki:2013wwa, Zurek:2013wia,DuttaBanik:2020vfr, Barman:2021ost, Cui:2020dly} and the other where BAU is generated from WIMP DM annihilations \cite{Yoshimura:1978ex, Barr:1979wb, Baldes:2014gca, Chu:2021qwk, Cui:2011ab, Bernal:2012gv, Bernal:2013bga, Kumar:2013uca, Racker:2014uga, Dasgupta:2016odo, Borah:2018uci, Borah:2019epq, Dasgupta:2019lha, Mahanta:2022gsi}. Other cogenesis scenarios motivated by the Affleck-Dine (AD) mechanism \cite{Affleck:1984fy} also exist in the literature \cite{Cheung:2011if, vonHarling:2012yn, Borah:2022qln}. 

In this letter, we propose a novel scenario where BAU is generated from DM decay. While DM is cosmologically stable, it can decay in the early universe when finite-temperature effects enable the forbidden decay modes. While the effects of forbidden decay on DM production have been discussed in the literature \cite{Darme:2019wpd, Konar:2021oye,Chakrabarty:2022bcn}, its role in cogenesis has not received any attention. In this work, we show that DM can decay during a finite period into SM leptons by virtue of finite-temperature effect generating a non-zero lepton asymmetry which later gets converted into baryon asymmetry by electroweak sphalerons. While this decay itself is not the source of asymmetry, it transfers part of the asymmetry in DM sector into the lepton sector. The DM sector asymmetry is generated by the AD mechanism. An AD field which explicitly breaks lepton number leads to a lepton asymmetry during cosmological evolution followed by its transfer to dark sector via decay. The same AD field also gives rise to non-minimal quartic inflation leading to the required inflationary parameters, as constrained by CMB data \cite{Akrami:2018odb, BICEP:2021xfz}. The requirement of successful cogenesis not only constrains the model parameters but also predicts a large self-interaction of DM which can have astrophysical implications \cite{Spergel:1999mh, Tulin:2017ara, Bullock:2017xww}. Therefore, the minimal setup with only four BSM fields capable of solving several cosmic puzzles remains verifiable in future cosmology, astrophysics as well as particle physics based experiments.

\vspace{0.5cm}
\noindent
{\bf The framework:} In order to realise the idea, we consider the four BSM fields as shown in table \ref{tab:Lepto1}. The scalar field $\Phi$ with non-zero lepton number plays the role of the AD field. The Dirac fermion $\chi$, stabilised by a $Z_2$ symmetry, plays the role of DM. The other two scalars $H_2, S$ assist in transferring the dark sector asymmetry partially to the lepton sector via forbidden decay of DM.

\begin{figure}
    \centering
    \includegraphics[scale=0.22]{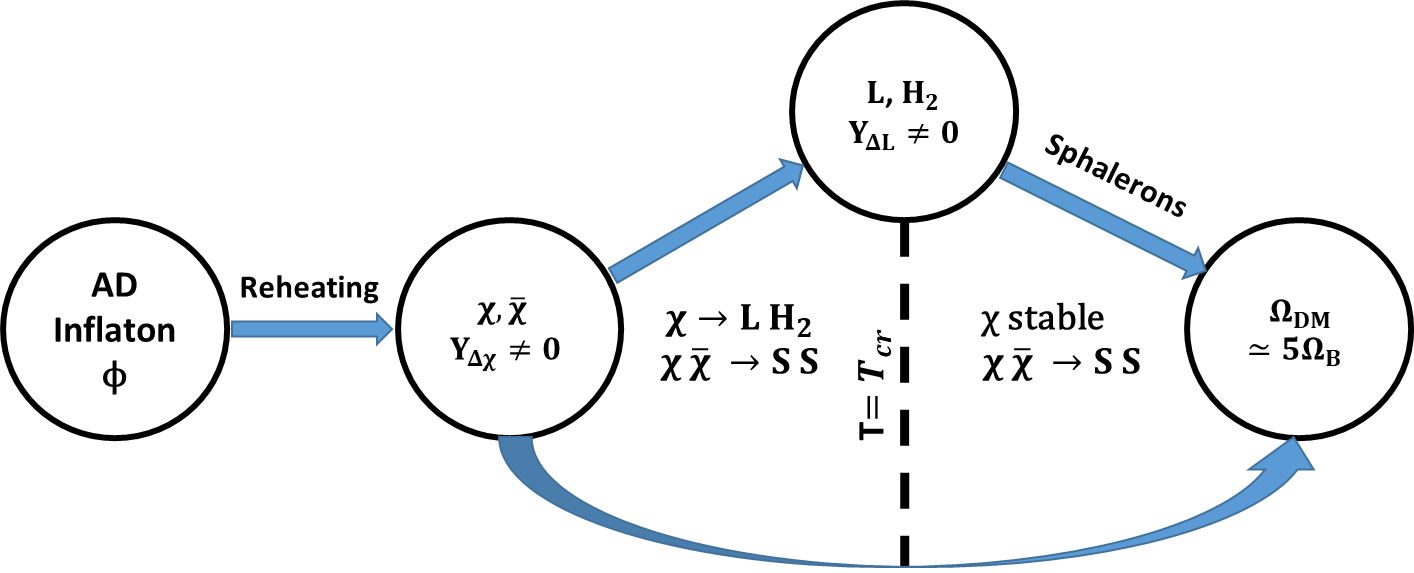}
    \caption{Schematic timeline of the cogenesis.}
    \label{fig:1}
\end{figure}

The relevant part of the Lagrangian is given by
\begin{align}
    -\mathcal{L} \supset   M_\chi \overline{\chi}\chi +Y_\nu \overline{L} \tilde{H_2} \chi_R + Y_D \overline{\chi^c} \chi \Phi^{\dagger} + Y_S\overline{\chi_L}\chi_R S+{\rm h.c.}
    \label{eq:L}
\end{align}
with $L$ being the SM lepton doublet. While these interactions conserve $U(1)_L$, the scalar potential of the AD field explicitly breaks it due to $\mu^2 \Phi^2$ term. The AD field also has non-minimal coupling to gravity $\mathcal{L}_{\rm inf} (\Phi,R)=-\frac{1}{2} \left( M_P^2 + \xi |\Phi|^2 \right)R$ which reproduces the successful inflationary cosmology \cite{Bezrukov:2007ep}. The cosmic evolution of $\Phi$ leads to a non-zero lepton asymmetry which then gets transferred to DM sector from $\Phi \rightarrow \chi \chi$ decay. The decay products namely, $\chi$ can reheat the universe instantaneously due to efficient annihilations. The forbidden decay of DM $\chi \rightarrow L H_2$ is allowed at high temperatures leading to partial transfer of dark sector asymmetry into leptons. At the same time, the symmetric part of DM annihilates away leaving only the asymmetric part. At a later epoch, the transfer of DM asymmetry to leptons via decay gets kinematically forbidden while the transfer via scattering remains negligible throughout due to the suitable choice of parameters. 
This is summarised in the schematic diagram shown in Fig. \ref{fig:1}. As shown in this schematic diagram, successful realisation of the idea in this particular setup relies upon the following criteria.
\begin{itemize}
    \item $\chi \rightarrow L H_2$ is forbidden below a critical temperature $T_{\rm cr}$. At $T>T_{\rm cr}$, this decay is allowed, transferring the dark asymmetry partially to lepton sector above sphaleron decoupling temperature. 
    \item If the interaction $Y_\nu \overline{L} \tilde{H_2} \chi_R$ is in equilibrium, then asymmetry can be transferred via scattering too, without relying on finite temperature effects required for decay. This requires Yukawa coupling $Y_\nu$ to be tiny such that asymmetry gets transferred dominantly via decay while keeping the scattering out-of-equilibrium throughout.
    \item For $T< T_{\rm cr}$, $H_2$ can start decaying into $\chi, L$. Since $H_2$ is complex, it can be asymmetric due to the production from $\chi$ at $T> T_{\rm cr}$. In order to ensure that late decay of $H_2$ does not washout the lepton asymmetry, $ H_2 \leftrightarrow H^\dagger_2$ type of interactions should be efficient at $T \sim T_{\rm cr}$. 
\end{itemize}

\begin{table}
    \centering
    \begin{tabular}{|c|c|c|c|}
    \hline
    Fields & $SU(3)_c \times SU(2)_L \times U(1)_Y$  & $U(1)_L$ & $Z_2$\\
    \hline  
      $ \chi$ & $(1,1,0)$ & 1 & -1 \\
        $H_2$ & $(1,2,-1/2)$ & 0 & -1 \\
          $ \Phi$ & $(1,1,0)$ & 2 & 1 \\
           $ S$ & $(1,1,0)$ & 0 & 1 \\
          \hline
    \end{tabular}
    \caption{BSM field content of the model.}
  \label{tab:Lepto1}
\end{table}

\vspace{0.5cm}
\noindent
{\bf Cogenesis of DM and baryon:} Before proceeding to calculate the abundances of DM and lepton asymmetry, we first find the finite-temperature masses of DM $\chi$, second Higgs doublet $H_2$, singlet scalar $S$ as well as lepton doublet $L$. The details are shown in appendix \ref{appen1}. 
The relevant parameters in Eq. \eqref{eq:L} are chosen in such a way that the desired mass spectrum of $\chi, H_2, L$ at $T > T_{\rm cr}$ as well as $T < T_{\rm cr}$ can be obtained.
While a strong coupling of $S$ to DM helps in generating a large thermal mass of $\chi$, a light $S$ can also help in annihilating away the symmetric part of DM via the $\chi \bar{\chi} \rightarrow S S$ process, in the spirit of cogenesis. We consider $S$ to be in equilibrium while writing the relevant Boltzmann equations for DM and leptons.
The relevant Boltzmann equations for $\chi, \bar{\chi}, L, \bar{L}$ are written in appendix \ref{appen2}.



\begin{figure}[htb!]
\centering
\includegraphics[scale=0.6]{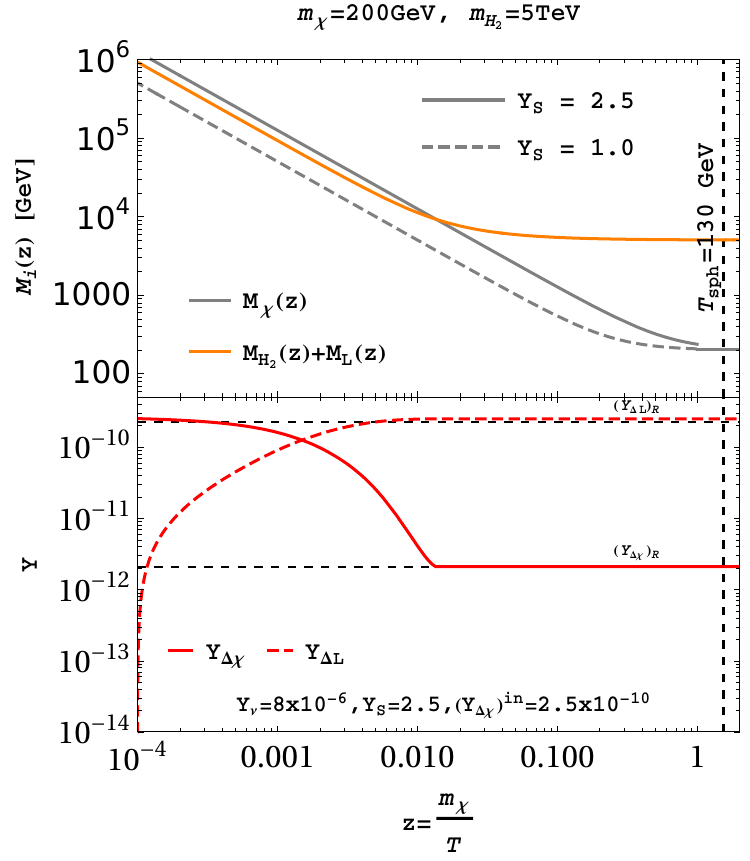}
\caption{Top panel: Variation of thermal masses of different particles with $z=\frac{m_\chi}{T}$ for two different values of $Y_S$. Bottom panel: Evolution of the comoving baryon and DM asymmetries with $z=\frac{m_{\chi}}{T}$ for a fixed value of initial dark asymmetry $(Y_{\Delta \chi})^{\text{in}}=2.5\times10^{-10}$. The vertical dashed line in both the panels corresponds to the sphaleron freeze-out temperature. The horizontal dashed lines in the bottom panel correspond to the required lepton asymmetry $(Y_{\Delta L})_{\text{R}}$ and the required dark sector asymmetry $(Y_{\Delta \chi})_{\text{R}}$ respectively. For both the panels we choose $m_{H_2}=5~\text{TeV}$, $m_{\chi}=200~\text{GeV}$.}
\label{thermal_mass}
\end{figure}

\begin{figure}[htb!]
\centering
\includegraphics[scale=0.6]{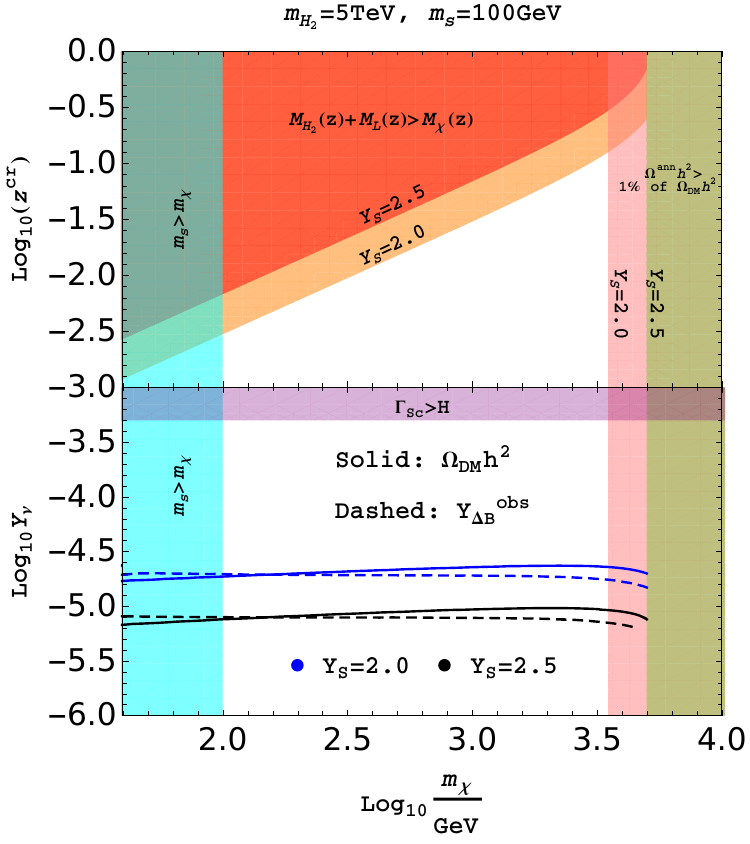}
\caption{Top panel: Variation of critical temperature equivalent $z^{\rm cr}=\frac{m_\chi}{T_{\rm cr}}$ with DM mass $m_\chi$ for two different values of $Y_S$.  Bottom panel: Contours consistent with correct baryon asymmetry and asymmetric DM relic in $Y_\nu-m_\chi$ plane for two different values of $Y_S$. Rightmost shaded region corresponds to the parameter space where symmetric component of DM contributes more than $1\%$ of total DM relic. Leftmost shaded region corresponds to inefficient DM annihilation due to $m_S > m_\chi$. For both the panels we have fixed $m_{H_2}=5~\text{TeV}$, $m_{S}=100~\text{GeV}$.}
\label{contour}
\end{figure}

\begin{figure}[htb!]
\centering
\includegraphics[scale=0.6]{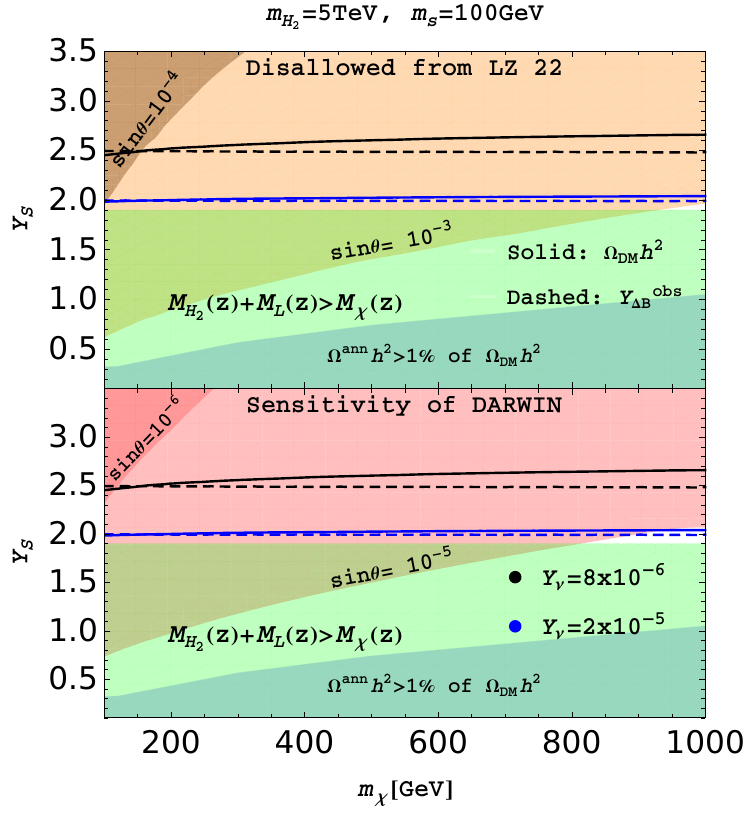}
\caption{Contours consistent with correct baryon asymmetry and asymmetric DM relic in $Y_S-m_\chi$ plane for two different values of $Y_\nu$. The shaded region (dark brown and light brown) in upper panel corresponds to the current experimental constraints from DM direct detection experiments LZ 2022 while the dark pink and light pink shaded regions in bottom panel correspond to future sensitivity, for different choices of scalar mixing. The green shaded region shows the parameter space where the DM remains always stable. For both the panels we have fixed $m_{H_2}=5~\text{TeV}$, $m_{S}=100~\text{GeV}$.}
\label{DDcontour}
\end{figure}

In the top panel of Fig.~\ref{thermal_mass}, we show the variation of the thermal mass of DM ($M_\chi(z)$) and the sum of the thermal masses of $H_2$ and the SM lepton doublet $i.e.~M_{H_2}(z)+M_L(z)$ with $z=m_\chi/T$. Here we set $M_\chi(T=0)=m_\chi=200$ GeV and $M_{H_2}(T=0)=m_{H_2}=5$ TeV and show the variations for two different values of $Y_S=2.5$ and $1.0$. While $M_{H_2}(z)+M_L(z)$ remains independent of $Y_S$ as expected, a clear dependence of $M_\chi (z)$ can be seen on $Y_S$. Note that, in order to generate the lepton asymmetry from the DM's forbidden decay, one needs to satisfy the condition: $M_\chi(z)>M_{H_2}(z)+M_L(z)$ at some stage in the early universe. From the top panel of Fig.~\ref{thermal_mass}, it is clear that this condition can be satisfied for an appropriate choice of $Y_S$. We also define a critical value ($z^\text{cr}$) of $z$ at which $M_\chi(z^\text{cr})=M_{H_2}(z^\text{cr})+M_L(z^\text{cr})$ is satisfied. In other words, successful leptogenesis through the forbidden decay of DM can only be achieved in a region where $z<z^\text{cr}$. For $z>z^\text{cr}$, the production of lepton asymmetry stops. For $Y_S=2.5$, $z^\text{cr}=0.14.$ In this figure, we do not have any critical values of $z$ with $Y_S=1.0$ as the condition $M_\chi(z)>M_{H_2}(z)+M_L(z)$ is never achieved. Since the sphaleron decoupling occurs around a temperature $T_\text{sph}\simeq130$ GeV ($z_\text{sph}$), any lepton asymmetry produced at $z > z_\text{sph}$ is not converted into the baryon asymmetry.

In the bottom panel of Fig.~\ref{thermal_mass}, we show the evolution of the dark sector asymmetry (solid) together with the baryon asymmetry (dotted) with $z=m_\chi/T$ obtained after solving the set of coupled Boltzmann equations involving DM and leptons. The comoving asymmetry for a species $x$ is defined as $Y_{\Delta x}=(n_x-n_{\bar{x}})/s$ with $n_x, s$ being the number density of species $x$ and entropy density of the universe respectively. As a result of the decay of the AD field ($\Phi$) to the DM ($\chi$), a lepton asymmetry is generated among the DM particle and its antiparticle. If kinematically allowed, DM can further decay to the $H_2, L$ by virtue of finite-temperature effects while transferring its asymmetry to the lepton sector that can be further converted to baryon asymmetry ($Y_{\Delta B}$) via electroweak sphalerons. 
Here for the first time, we show a decay of the DM through its forbidden channel can generate the visible sector asymmetry without affecting its stability condition at the present universe. We first set the initial asymmetry in the dark matter produced from the decay of AD field at $Y_{\Delta\chi}^{\text{int}}=2.5\times10^{-10}$. 
As a result of the forbidden DM decay, the dark sector asymmetry is partially transferred to the lepton sector and hence a rise is observed in the yield of lepton asymmetry $Y_{\Delta L}$ whereas an equivalent fall is observed in the dark sector asymmetry. This increasing (decreasing) behaviour of $Y_{\Delta L}$ ($Y_{\Delta \chi}$) stops when the threshold $M_\chi(z^\text{cr})=M_{H_2}(z^\text{cr})+M_L(z^\text{cr})$ is hit. Thereafter, the asymmetries in both sectors saturate. We find that for $Y_\nu=2.5\times10^{-6}$, the observed baryon asymmetry of the universe ($Y_{\Delta B}^{\text{obs}}=8.75\times 10^{-11}$~\cite{Aghanim:2018eyx}) together with observed DM relic abundance ($\Omega_\text{DM}h^2=0.12$~\cite{Aghanim:2018eyx}) with DM mass $m_\chi=200$ GeV can be explained.

In the top panel of Fig.~\ref{contour}, we show a region of parameter space (in white) in $z^\text{cr}-m_\chi$ plane where baryogenesis via leptogenesis can proceed through the forbidden decay of the DM. As observed earlier, a large $Y_S$ is required in order to have a successful leptogenesis through this forbidden channel. Such large values of $Y_S$ also help in the rapid annihilation of DM to get rid of the symmetric part, a requirement in typical asymmetric DM scenarios. 
While DM has Yukawa interactions with leptons, the corresponding coupling $Y_\nu$ is required to be small for reasons discussed below. We show the region of parameter space (in pink and green) corresponding to $\Omega^\text{ann}h^2>1\%~\text{of}~\Omega_\text{DM}h^2$ in Fig.~\ref{contour} implying the symmetric part of DM contributing more than $1\%$ of the total DM relic and hence disfavoured in the spirit of asymmetric DM. As expected, due to its rapid annihilation to a lighter scalar (here we have considered $m_s=100$ GeV) even for satisfying $\Omega^\text{ann}h^2>1\%~\text{of}~\Omega_\text{DM}h^2$ a heavy DM with $m_\chi\gtrsim 3.5 (5)$ TeV is required for the $Y_S=2.0 (2.5)$. The cyan shaded region towards the left is disfavoured as $m_S >m_\chi$ will forbid efficient DM annihilation into light scalars at low temperatures. The shaded region in the upper left part denotes the region where DM is always stable. Since DM mass receives a larger thermal correction for larger $Y_S$, the critical temperature turns out to be smaller (or larger $z^{\rm cr})$, as evident from this plot.


In the bottom panel of Fig.~\ref{contour}, we show the contours satisfying correct baryon asymmetry (dashed line) and asymmetric DM relic (solid line) in $Y_\nu-m_\chi$ plane for two different choices of $Y_S$. The point at which these two contours intersect corresponds to successful cogenesis. The region corresponding to large $Y_\nu$ is disfavoured as it will bring scattering processes capable of transferring DM asymmetry to lepton into equilibrium leading to lesser dependence on forbidden decay of DM. It should be noted that the parameter space shown in the above figure satisfies the criteria $T_{\rm cr} >m_{H_2}$ which ensures that $H_2$ can be in equilibrium at $T=T_{\rm cr}$ with efficient conversions $H_2 \leftrightarrow H^\dagger_2$. Such conversions can occur independently of the parameters relevant for cogenesis and ensure that $H_2$ decay at $T< T_{\rm cr}$ does not wash out the lepton asymmetry generated at $T > T_{\rm cr}$ from forbidden DM decay. It should also be noted that, we have remained agnostic about the origin of light neutrino masses in our setup. The AD field breaks lepton number by $\Delta L=4$ units due to the $\mu^2\Phi^2$ term. Also, this field does not acquire any vacuum expectation value. Therefore, there is no source of generating Majorana mass of either $\chi$ or neutrinos in this minimal setup due to the absence of lepton number violation by $\Delta L=2$ units. In other words, our setup will work even if we have purely Dirac active neutrinos. On the other hand, the AD field itself can lead to washout of asymmetries and it is preferable to have $m_\Phi > T_{\rm RH}$. We have checked that for suitable choices of explicit lepton number violation by AD field and its coupling to DM namely $Y_D$, we can satisfy the required initial dark asymmetry while keeping the AD field out of equilibrium after reheating. Ensuring $m_\Phi > T_{\rm RH}$ also keeps the $\Delta L=4$ washouts like $\chi \chi \rightarrow \bar{\chi} \bar{\chi}$ suppressed. The details of dark sector asymmetry and washouts are given in appendix \ref{appen3}.

\vspace{0.5cm}
\noindent
{\bf Detection Prospects:} There are several promising detection prospects of the model we have proposed here. DM can scatter of nucleons due to singlet ($S$) mixing with the SM Higgs ($h$) leading to spin-independent DM-nucleon scattering tightly constrained by direct detection experiments like LZ \cite{LZ:2022ufs}. In Fig. \ref{DDcontour}, we show the current LZ limit and future sensitivity of DARWIN \cite{DARWIN:2016hyl} for different choices of singlet-SM Higgs mixing $\theta$ in $Y_S-m_\chi$ plane. The contours for chosen $Y_\nu$ indicate the cogenesis preferred parameter space. 
The green shaded regions corresponding to smaller values of $Y_S$ indicate the parameter space where forbidden DM decay is never allowed. For even smaller values of $Y_S$, the annihilation of DM is not sufficient enough to keep the symmetric part below $1\%$ of total DM relic. 

The model also has cosmological predictions due to role of $\Phi$ as inflaton via non-minimal coupling $(\xi)$ to gravity. When $\Phi> M_P/\sqrt{\xi}$, it slow-rolls and causes inflation generating the required tensor-to-scalar ratio and scalar spectral index \cite{Okada:2010jf, Okada:2015lia, Borah:2020wyc, Borah:2021inn} consistent with cosmological data from CMB experiments like Planck \cite{Akrami:2018odb} and BICEP/Keck \cite{BICEP:2021xfz}. For example, with $\xi\gg 1$, we have predictions for inflationary observables, namely the magnitude of spectral index ($n_s$) and tensor-to-scalar ratio ($r$) as $r=0.003$, $n_{s}=0.967$ for number of e-folds $N_e=60$, which satisfies Planck 2018 data at 1$\sigma$ level \cite{Akrami:2018odb}. 

Due to the strong coupling of DM with the singlet scalar, it is possible to have large self-interactions, having the potential to solve the small scale issues of cold DM like too-big-to-fail, missing satellite and core-cusp problems faced by the latter \cite{Spergel:1999mh, Tulin:2017ara, Bullock:2017xww}. For a light mediator, it is possible to have velocity dependent DM self-interactions in order to solve the small scale issues while being consistent with standard CDM properties at large scales \cite{Buckley:2009in, Feng:2009hw, Loeb:2010gj, Bringmann:2016din, Kaplinghat:2015aga, vandenAarssen:2012vpm, Tulin:2013teo}. For $m_S \ll m_\chi$, we can satisfy the required velocity-dependent self-interactions in our setup (similar to \cite{Borah:2021qmi} where fermion DM with light scalar mediator was studied), which can be probed via astrophysical observations at different scales, such as dwarfs, low surface brightness galaxies and clusters~\cite{Kaplinghat:2015aga,Kamada:2020buc}. 

Collider prospects of the model can be in terms of invisible SM Higgs decay into light scalar $S$ \cite{ATLAS:2020cjb} or signatures of heavy Higgs $H_2$. If produced in the large hadron collider (LHC), components of $H_2$ can lead to same-sign dilepton plus missing energy \cite{Gustafsson:2012aj, Datta:2016nfz}, dijet plus missing energy \cite{Poulose:2016lvz}, tri-lepton plus missing energy \cite{Miao:2010rg} or even mono jet signatures \cite{Belyaev:2016lok, Belyaev:2018ext}. Depending upon $hSS$ coupling, the Higgs invisible decay rate can saturate the current limit \cite{ATLAS:2020cjb}. The model can also have complementary detection prospects like gravitational waves (GW). As discussed in \cite{Zhou:2015yfa, White:2021hwi}, the fragmentation of the Affleck-Dine condensate can either generate GW or amplify primordial GW bringing it within sensitivities of ongoing and near future experiments.

\vspace{0.5cm}
\noindent
{\bf Conclusion:}
We have proposed a novel scenario where baryon asymmetry via leptogenesis occurs due to forbidden decay of dark matter. Dark matter acquires an asymmetry from an Affleck-Dine field which also plays the role of inflaton. Forbidden decay of DM into lepton and a second Higgs doublet, enabled by finite-temperature effects, leads to transfer of some dark sector asymmetry into leptons with the latter being converted into baryon asymmetry via electroweak sphalerons. The required finite temperature correction to DM mass can be obtained by virtue of its strong coupling to a singlet scalar. The same singlet scalar can also assist in annihilating away the symmetric component of DM in the spirit of asymmetric DM. While being consistent with correct baryon asymmetry and DM relic, the proposed setup can have a variety of detection prospects in terms of inflationary observables via CMB measurements, DM direct detection, DM self-interactions via light scalar as well as collider signatures of new scalars. These complementary detection prospects via cosmology, astrophysics and laboratory based observations keep this framework verifiable in near future.

\acknowledgements
The work of D.B. is supported by the Science and Engineering Research Board (SERB), Government of India grant MTR/2022/000575. R.R. acknowledges the National Research Foundation of Korea (NRF) grant
funded by the Korean government (2022R1A5A1030700) and also the support provided by the Department of Physics, Kyungpook National University.

\appendix
\section{Thermal masses}
\label{appen1}
In the proposed setup, we have four new fields beyond the standard model namely, the Affleck-Dine (AD) inflaton field $\Phi$, Dirac fermion dark matter $\chi$, an inert Higgs doublet $H_2$ and a scalar singlet $S$. The finite-temperature masses of relevant particles involved in forbidden decay are given by~\cite{Giudice:2003jh} 

\bea
M_\chi(T)&=&\sqrt{m_\chi^2+\Pi_{S\chi}^2(T)},
\label{chi_thermal_mass}\\
M_{H_2}(T)&=&\sqrt{m_{H_2}^2+\Pi_\text{gauge}^2(T)},
\label{H2_thermal_mass}\\
M_L(T)&=&\sqrt{m_L^2+\frac{1}{2}\Pi_\text{gauge}^2(T)},
\label{s_thermal_mass}
\eea
where
\bea
 \Pi_{S\chi}^2(T)&=&\frac{Y_{S}^2}{16}T^2,\\
 \Pi_\text{gauge}^2(T)&=&\bigg{(}\frac{1}{16}g^{\prime 2}+\frac{3}{16}g^{2}\bigg{)}T^2.
 \label{S_corrections}
\eea

\section{Boltzmann equations}
\label{appen2}
The relevant Boltzmann equations for generating lepton asymmetry from an initial dark sector asymmetry can be written as follows.
\begin{widetext}
\begin{equation}
 \frac{dY_{\chi}}{dz} =-\frac{s}{{\bf H}z}\big[\langle \sigma v_{\chi\bar{\chi}\rightarrow \rm SM}\rangle (Y_\chi Y_{\bar{\chi}}-Y_\chi^{\rm eq}Y_{\bar{\chi}}^{\rm eq})\big] -\frac{1}{s{\bf H}z}\gamma(\chi\to L H_2) \bigg(\frac{Y_\chi}{Y_\chi^{\rm eq}}-1\bigg)+\frac{1}{s{\bf H}z}\gamma(H_2\to \chi \bar{L}),    
\end{equation}
\begin{equation}
   \frac{dY_{\bar{\chi}}}{dz} =-\frac{s}{{\bf H}z}\big[\langle \sigma v_{\chi\bar{\chi}\rightarrow \rm SM}\rangle (Y_\chi Y_{\bar{\chi}}-Y_\chi^{\rm eq}Y_{\bar{\chi}}^{\rm eq})\big] -\frac{1}{s{\bf H}z} \gamma (\bar{\chi} \to \bar{L} H_2) \bigg( \frac{Y_{\bar{\chi}}}{Y^{\rm eq}_{\bar{\chi}}}-1 \bigg ) + \frac{1}{s{\bf H}z}\gamma(H_2\to \bar{\chi} L), 
\end{equation}
\begin{equation}
   \frac{dY_{L}}{dz}=\frac{1}{s{\bf H}z}\gamma(\chi\to L H_2) \bigg(\frac{Y_\chi}{Y_\chi^{\rm eq}}-1\bigg) + \frac{1}{s{\bf H}z}\gamma(H_2\to \bar{\chi} L),
\end{equation}
\begin{equation}
\frac{dY_{\bar{L}}}{dz} = \frac{1}{s{\bf H}z}\gamma(\bar{\chi}\to \bar{L} H_2) \bigg(\frac{Y_{\bar{\chi}}}{Y_{\bar{\chi}}^{\rm eq}}-1\bigg) + \frac{1}{s{\bf H}z}\gamma(H_2\to \chi \bar{L}),
\label{BE}
\end{equation}
\end{widetext}

\noindent
where $Y_i=n_i/s$ denotes comoving number density of species ``i" with $s$ being the entropy density. Hubble expansion rate is denoted by ${\bf H}$ while the variable $z$ is $m_\chi/T$. The reaction density $\gamma$ is given by
\bea
\gamma(a \rightarrow bc) = n^{\rm eq}\frac{K_1(z)}{K_2(z)}\Gamma(a \rightarrow bc),
\eea
where $K_{1,2}$ are Bessel functions of 1st, 2nd kind respectively and the decay width of $\chi\to L,H_2$ and $\bar{\chi}\to \bar{L} H_2$ are given by

\bea
\Gamma(\chi \rightarrow L H_2)&=&\Gamma(\bar{\chi}\to \bar{L} H_2)\nonumber\\&=& \frac{Y_\nu^2}{16\pi}M_{\chi}\bigg(1-\frac{(M_{H_2}+M_L)^2}{M_\chi^2}\bigg)^{1/2} \nonumber \\
& \times & \bigg(1-\frac{(M_{H_2}-M_L)^2}{M_\chi^2}\bigg)^{1/2} \nonumber \\
& \times & \bigg(1-\frac{(M_{H_2}^2-M_L^2)}{M_\chi^2}\bigg).
\label{decay_chi}
\eea
Note that we have treated $H_2$ and $H^\dagger_2$ on equal footing under the assumption that any asymmetry in $H_2$ can be washed out due to $H_2 \leftrightarrow H^\dagger_2$ conversions. Due to the possibility of scalar portal interactions with the SM Higgs doublet, such conversions can occur independently of the interactions relevant for the above equations. 

Next we define $Y_{\Delta \chi}=Y_{\chi}-Y_{\bar{\chi}}$ and $Y_{\Delta L}=Y_{L}-Y_{\bar{L}}$. We choose the following initial condition for solving the above coupled  Boltzmann equations
\bea
Y_{\chi}(0)=Y_{\chi}^{\rm eq},~ Y_{\bar{\chi}}(0)=Y_{\chi}^{\rm eq}-Y^{\rm in}_{\Delta \chi}\\
Y_{L}(0)=Y_{L}^{\rm eq},~Y_{\bar{L}}(0)=Y_{L}^{\rm eq}
\eea
The initial dark sector asymmetry $Y^{\rm in}_{\Delta \chi}$ in required amount can be generated from the Affleck-Dine (AD) field as we discuss in \ref{appen3}.

\begin{figure}[htb!]
\centering
\includegraphics[scale=0.6]{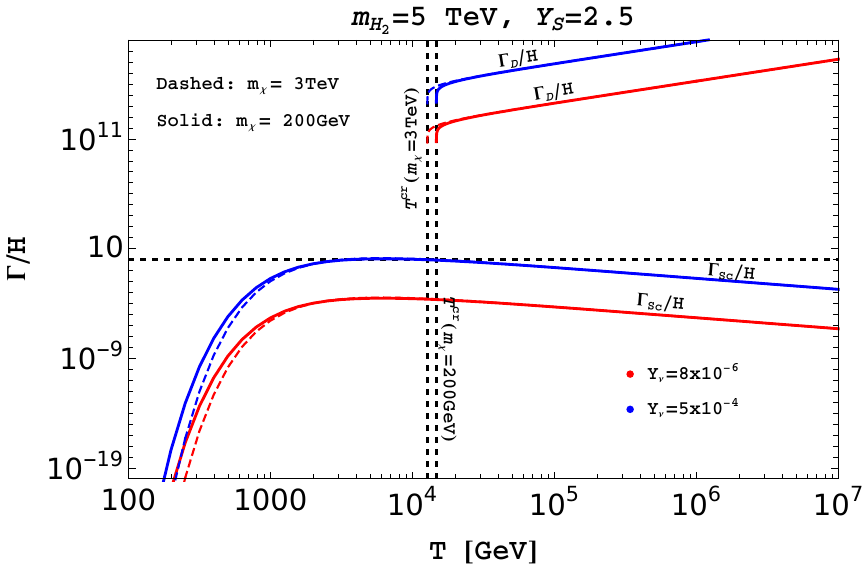}
\caption{Comparison of decay and scattering rates responsible for transferring dark sector asymmetry into leptons.}
\label{rate}
\end{figure}

While we have considered only the decays involving $\chi, H_2, L$ in the Boltzmann equations, responsible for transferring the dark sector asymmetry into leptons, it is also possible to have scatterings like $\chi \, H_2  \rightarrow L \, X$ transferring the asymmetry with $X$ being one of the allowed SM scalar/vector bosons present in the bath. In Fig. \ref{rate}, we show the comparisons of these decay and scattering rates. While for $T > T_{\rm cr}$, decay dominates over scattering significantly, for $T< T_{\rm cr}$, where decay is forbidden, the scattering rate also remains suppressed. For the chosen values of $Y_\nu$, the scattering remains out-of-equilibrium throughout validating the production of lepton asymmetry dominantly from decay.

\begin{figure}[htb!]
\centering
\includegraphics[scale=0.6]{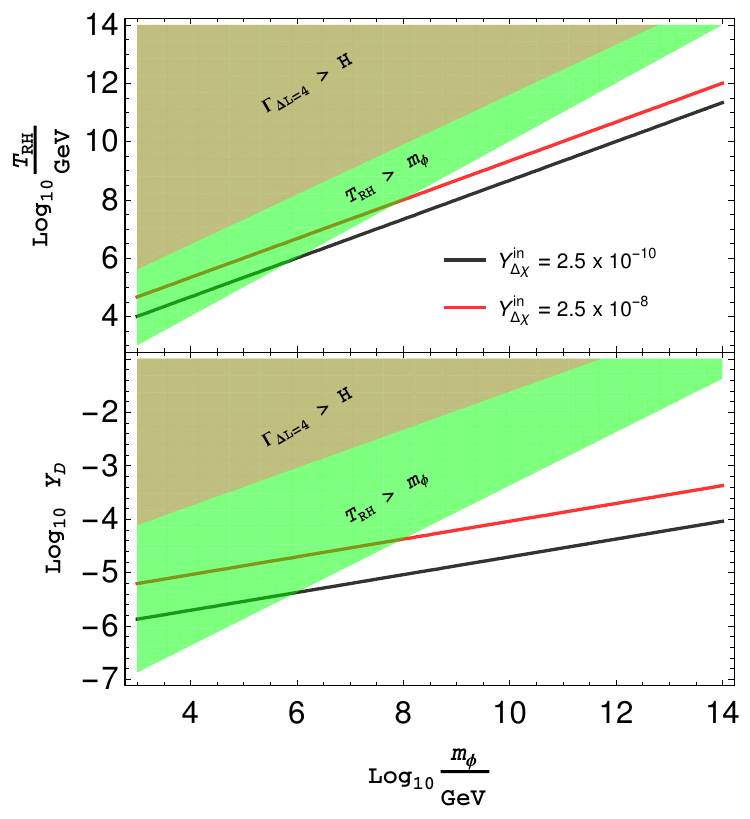}
\caption{Contours of $Y_{\Delta\chi}^{\text{in}}$ in the  $m_{\Phi}$-$T_{\text {RH}}$ plane (top) and $m_{\Phi}$-$Y_D$ plane (bottom). Here, we consider $\epsilon = 1.65 \times 10^{-3}$.}
\label{fig:AD}
\end{figure}

\section{Dark Asymmetry from Affleck-Dine field}
\label{appen3}
Since the AD field $\Phi$ carries a non-zero lepton number, a term in the scalar potential $\epsilon \mu^2 \Phi^2$ breaks the lepton number symmetry explicitly while all other terms conserve it. Due to this explicit lepton number violating term, the cosmic evolution of $\Phi$ leads to a net lepton asymmetry which gets transferred to the dark sector. The same decay of AD inflaton field to dark matter also reheats the universe to a temperature $T_{\rm RH}$. The asymmetry initially rises from zero and then oscillates until $t\gtrsim 1/\Gamma_{\Phi}$, when its amplitude exponentially damps to reach the constant value given by \cite{Lloyd-Stubbs:2020sed,Mohapatra:2021aig, Borah:2022qln}
\begin{equation}
Y^{\rm in}_{\Delta \chi} = \frac{(n_{\chi}-n_{\bar{\chi}})^{\rm in}}{s}\simeq \frac{T_{\rm RH}^{3}}{\epsilon m_{\Phi}^{2}M_{P}}.\label{eq:vis}
\end{equation}
As the decay $\Phi \rightarrow \chi \chi$ also reheats the universe, the reheating temperature is $T_{\rm RH}\simeq\sqrt{\Gamma_{\Phi}M_{P}}$ with $\Gamma_\Phi$ being the corresponding decay width. Now, the presence of lepton number violating interaction given by $\epsilon$ can lead to the washout of the generated asymmetry. This can happen through scatterings with  $\Delta L=4$ : $\chi \chi \leftrightarrow \overline{\chi}~\overline{\chi}$, mediated by $\Phi$ exchange and the $\epsilon$ term. If the decoupling temperature of such process is higher than the reheat temperature $T_{\rm RH}$, the washout effect would be absent. This leads to the following condition
\begin{equation}
T_{\rm RH}^{3}\frac{Y_{D}^{4}\epsilon^2T_{\rm RH}^2}{4 \pi m_{\Phi}^{4}}\lesssim \sqrt{\frac{\pi^2}{90}g_*} \frac{T_{\rm RH}^{2}}{M_{P}},\label{eq:washout}
\end{equation}
where $Y_D$ is the coupling of AD field to DM. In Fig. \ref{fig:AD}, we show contours of constant $Y_{\Delta\chi}^{\text{in}}$ in the $m_{\Phi}$-$T_{\text {RH}}$ (top panel) and $m_{\Phi}$-$Y_D$ plane (bottom panel). In the green-shaded region, $T_{\text {RH}} > m_{\Phi}$, which can lead to a washout of the asymmetry and hence disfavored. In the brown-shaded region, $\Delta L=4$ scatterings (with interaction rate denoted by $\Gamma_{\Delta L = 4}$) of the form $\chi \chi \leftrightarrow \overline{\chi}~\overline{\chi}$, mediated by $\Phi$ exchange can lead to washout of the asymmetry. This clearly justifies the choice of initial dark asymmetry considered in solving the Boltzmann equations.


\end{document}